\documentclass[prd,nofootinbib,preprint,superscriptaddress,twocolumn,10pt]{revtex4}
\pdfoutput=1
\usepackage[T1]{fontenc}
\usepackage{amsmath,amssymb}
\usepackage{braket}
\usepackage{epsfig}
\usepackage{graphicx}
\usepackage[usenames,dvipsnames]{color}
\usepackage{subfigure, orcidlink}
\usepackage{slashed}
\usepackage{enumerate}
\usepackage{placeins}

\usepackage{pdfpages}
\usepackage{color}
\usepackage{comment}
\usepackage{tikz-feynman}

\usepackage{amsfonts}
\DeclareMathSymbol{\shortminus}{\mathbin}{AMSa}{"39}
\begin{document}
\title{Matter-Antimatter Asymmetry from Dirac-Majorana Flip-Flop}

\author{Debasish Borah \orcidlink{0000-0001-8375-282X}}
\email{dborah@iitg.ac.in}
\affiliation{Department of Physics, Indian Institute of Technology Guwahati, Assam 781039, India}

\author{Arnab Dasgupta \orcidlink{0000-0003-1065-7279}}
\email{arnabdasgupta@pitt.edu}
\affiliation{Pittsburgh Particle Physics, Astrophysics, and Cosmology Center, Department of Physics and Astronomy, University of Pittsburgh, Pittsburgh, PA 15260, USA}

\author{Debalina Nandy \orcidlink{0009-0000-9164-7819}}
\email{n.debalina@iitg.ac.in}
\affiliation{Department of Physics, Indian Institute of Technology Guwahati, Assam 781039, India}

\author{Indrajit Saha \orcidlink{0000-0002-7459-0838}}
\email{s.indrajit@iitg.ac.in}
\affiliation{Department of Physics, Indian Institute of Technology Guwahati, Assam 781039, India}

\author{Harman Singh \orcidlink{0009-0000-0113-3276}}
\email{hps25@pitt.edu}
\affiliation{Pittsburgh Particle Physics, Astrophysics, and Cosmology Center, Department of Physics and Astronomy, University of Pittsburgh, Pittsburgh, PA 15260, USA}

\preprint{PITT-PACC-2616}

\begin{abstract}
The observed matter-antimatter asymmetry can be generated dynamically via leptogenesis, where a non-zero lepton asymmetry is generated first and then gets converted into baryon asymmetry via sphalerons. Creation of non-zero lepton asymmetry typically requires lepton number violation (LNV) which, within the seesaw framework, results in the Majorana nature of light neutrinos. With no signs of LNV at terrestrial experiments, we propose a novel leptogenesis scenario where LNV occurs at high temperatures while zero-temperature physics is consistent with the pure Dirac nature of neutrinos. In a concrete model of this "Dirac-Majorana flip-flop", the right chiral parts of Dirac neutrinos initially acquire a large Majorana mass in the early Universe when a singlet scalar field with two units of lepton number acquires a non-zero vacuum expectation value (VEV), thereby facilitating leptogenesis from out-of-equilibrium decay of heavy Majorana fermions. At lower temperatures, the VEV of the scalar singlet gets restored to zero to be consistent with lepton number conservation. A first-order VEV-restoring phase transition can lead to observable gravitational waves, while light Dirac neutrinos can generate detectable dark radiation at cosmic microwave background experiments. 
\end{abstract}
\maketitle
\noindent
\textbf{\textit{Introduction:}} The visible or baryonic matter content of the present Universe is highly asymmetric, which leads to a puzzle known as the baryon asymmetry of the Universe (BAU)~\cite{ParticleDataGroup:2026mpi, Planck:2018vyg, Bodeker:2020ghk}. The observed BAU is quantified in terms of the baryon-to-photon number density ratio as~\cite{Planck:2018vyg} 
\begin{equation}
\eta_B = \frac{n_{B}-n_{\overline{B}}}{n_{\gamma}} = (6.12\pm 0.04) \times 10^{-10}, 
\label{etaBobs}
\end{equation} 
based on cosmic microwave background (CMB) measurements, which is consistent with the value extracted from big bang nucleosynthesis (BBN) observations of primordial elemental abundances~\cite{ParticleDataGroup:2026mpi}. While the standard model (SM) has, in principle, all the basic ingredients to satisfy the Sakharov conditions needed to dynamically generate baryon asymmetry starting from a symmetric Universe~\cite{Sakharov:1967dj, Cline:2006ts}, the quark sector CP violation falls short by almost 10 orders of magnitude~\cite{Shaposhnikov:1987pf}. In addition, the electroweak phase transition in the SM is not strongly first-order enough to ensure adequate departure from equilibrium, necessary to produce the BAU~\cite{Kajantie:1996mn}. This has led to several beyond standard model (BSM) proposals, out of which baryogenesis~\cite{Weinberg:1979bt, Kolb:1979qa} or leptogenesis~\cite{Fukugita:1986hr} have been the most widely studied ones. A typical baryogenesis scenario involves out-of-equilibrium decay of heavy new particles. In leptogenesis scenarios, a non-zero lepton asymmetry is first generated, which later gets converted into the BAU via electroweak sphalerons~\cite{Kuzmin:1985mm}. An appealing feature of leptogenesis is its natural realization within the seesaw mechanism~\cite{Minkowski:1977sc,Yanagida:1979as,Gell-Mann:1979ijt,GellMann:1980vs,Mohapatra:1979ia,Schechter:1980gr,Schechter:1981cv} for which explains the origin of neutrino mass and mixing, another experimental observation~\cite{ParticleDataGroup:2026mpi} which the SM fails to address.

Typical leptogenesis scenarios rely on lepton number violation (LNV) resulting in the Majorana nature of light neutrinos. However, it is possible to have successful leptogenesis with light Dirac neutrinos, where the total lepton number or $B-L$ number is conserved. In these Dirac leptogenesis scenarios, an equal and opposite amount of lepton asymmetry is generated in the left-handed and right-handed neutrino sectors, followed by the conversion of the left sector asymmetry into baryon asymmetry via electroweak sphalerons. The left sector asymmetry survives until sphaleron decoupling as it is prevented from equilibrating with the right sector due to the tiny effective Dirac Yukawa couplings. Various possible implementations of this idea can be found in, for example, Refs.~\cite{Boz:2004ga, Thomas:2005rs, Thomas:2006gr, Cerdeno:2006ha, Gu:2006dc, Gu:2007mc, Chun:2008pg, Bechinger:2009qk, Chen:2011sb, Choi:2012ba, Borah:2016zbd, Gu:2016hxh, Narendra:2017uxl, Barman:2022yos, Barman:2023fad, Borboruah:2024lli, Blazek:2024efd, Borah:2025dka, Ahmed:2025vzl, Borah:2026vqw, Manno:2026mws}. In a few related works~\cite{Heeck:2013vha, Gu:2019yvw, Mahanta:2021plx}, a net violation of $B-L$ symmetry was accommodated in a way to preserve the Dirac nature of light neutrinos, while generating lepton asymmetry simultaneously. 

In this letter, we provide a novel alternative by realizing conventional leptogenesis with LNV in a light Dirac neutrino setup by considering a flip-flop of Dirac and Majorana nature of neutrinos in the early Universe. This relies on lepton number violation in the early Universe due to finite-temperature effects whereas the observations in the present Universe remain consistent with global lepton number conservation. While gauged or global symmetry non-restoration in the early Universe is studied in several works \cite{Weinberg:1974hy, Mohapatra:1979qt, Salomonson:1984rh, Fujimoto:1984hr, Bimonte:1995sc, Dvali:1995cj, Dvali:1996zr, Gavela:1998ux, Cline:1999wi, Espinosa:2004pn, Patel:2013zla, Meade:2018saz, Baldes:2018nel, Matsedonskyi:2020mlz, Matsedonskyi:2021hti, Carena:2021onl, Bai:2021hfb, Chao:2021xqv, Borah:2026aqv}, the consequences for baryogenesis or leptogenesis has not received much attention \cite{Dodelson:1990ge, Sakstein:2017lfm, Dine:1989kt, Lopez-Zurdo:2023xmr}. Here we consider the non-restoration of lepton number symmetry in the early Universe at the perturbative level due to a singlet scalar with non-zero lepton number acquiring a non-zero vacuum expectation value (VEV). The same VEV also generates Majorana masses for right-handed neutrinos (RHNs), followed by the creation of lepton asymmetry from out-of-equilibrium decay of RHNs into a lepton and a neutrinophilic Higgs doublet. At a later epoch, the singlet scalar VEV shifts to zero, restoring lepton number symmetry, whereas the lepton asymmetry generated gets converted into baryon asymmetry via sphalerons. The VEV restoring phase transition, responsible for Dirac-Majorana flip-flop, can be first-order, with additional detection prospects like stochastic gravitational waves (GWs). A sizeable Yukawa coupling of Dirac neutrinos to the neutrinophilic Higgs doublet not only assists in leptogenesis, but also leads to the thermalization of the right-chiral parts, enhancing dark radiation or $\Delta N_{\rm eff}$, which can be observed at CMB experiments. \\

\begin{figure}[h]
    \centering
    \includegraphics[width=0.99\linewidth]{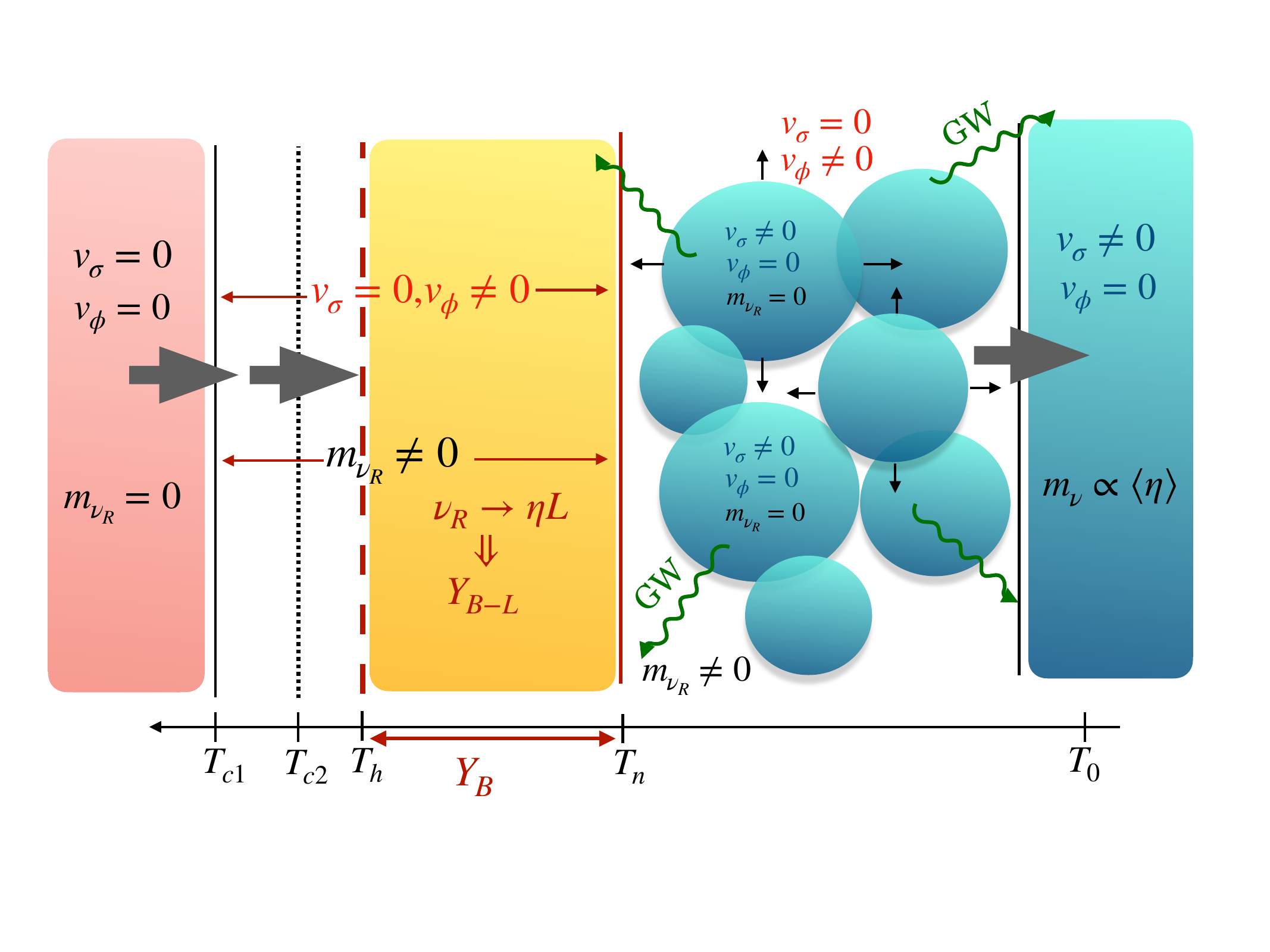}
    \caption{ The schematic of the proposed Dirac-Majorana flip-flop scenario. At a temperature $T>T_{c1}$ the Universe remains in the symmetric phase $(\sigma, \phi) \equiv (0,0)$ (pink region) and undergoes a transition (second-order/crossover) to $(0,v_{\phi})$ at $T_{c1}$. This allows the RHNs to acquire mass through the VEV of $\phi$. As the Universe further cools down, a FOPT occurs at $T_{c2}<T_{c1}$ from $(0,v_{\phi})$ to $(v_{\sigma},0)$. Bubbles of the new vacuum nucleate at $T_n<T_{c2}$. Hence, the RHNs remain massive in the old vacuum during $T_{n}<T<T_{c2}$. Leptogenesis starts only below $T_h$ when the $\nu_R\rightarrow \eta L$ decay becomes kinematically allowed and continues till the RHNs are massive, i.e up to $T=T_n$ (yellow region). Below $T_n$, the Universe gradually gets filled with the true vacuum $(v_{\sigma},0)$ (blue region), allowing the Majorana mass of RHNs to vanish. At present temperature, the neutrinos develop a sub-eV Dirac mass from the small VEV of the neutral component of $\eta$ after electroweak symmetry breaking.}
    \label{Fig:Schematic}
\end{figure}

\noindent
\textbf{\textit{The Framework:}} To implement our idea, we consider a toy model with two singlet scalars $\phi, \sigma$ undergoing a two-step phase transition. At $T=T_{c1}$, the scalar field $\phi$ acquires a non-zero VEV imparting Majorana masses to RHNs. This allows the generation of lepton asymmetry from out-of-equilibrium decay of RHNs. At a lower temperature $T=T_{c2}$, the VEV of $\phi$ gets restored to zero while $\sigma$ acquires a VEV. If the second step is a first-order phase transition (FOPT), the generation of lepton asymmetry continues in the false vacuum until the nucleating bubbles of true vacuum occupy the entire Hubble horizon. The conceptual visualization of the idea is shown as a schematic in Fig. \ref{Fig:Schematic}.

Apart from the two singlet scalars, the SM is extended by two copies of gauge singlet right-handed neutrinos $\nu_R$, one Majorana dark fermion singlet $\chi$, and a neutrinophilic Higgs doublet $\eta$. The relevant Yukawa Lagrangian is given by 
\begin{equation}
\label{eq:YukLag}
    -\mathcal{L}_Y \supset y_{\alpha i} \overline{L}_\alpha \tilde{\eta} \nu_{R_i} + Y_{ij} \phi \overline{\nu^c_{R_i}} \nu_{R_j} + Y_\chi \sigma \overline{\chi^c}\chi + {\rm h.c.}
\end{equation}
At high temperatures $T > T_{\rm sph} \sim 131.7 $ GeV, $\phi$ acquires a large VEV $v_\phi \gg T_{\rm sph}$ generating Majorana mass terms for $\nu_R$. This allows $\nu_R \rightarrow L \eta$ decay to generate non-zero lepton asymmetry, which gets converted into baryon asymmetry during the sphaleron decoupling epoch $T = T_{\rm sph}$. A softly broken $Z_2$ symmetry under which $\eta, \nu_R$ are odd ensures the validity of the Yukawa interactions in Eq. \eqref{eq:YukLag}. While the inclusion of $\nu_R, \eta$ is related to the origin of neutrino mass and leptogenesis, the dark fermion $\chi$ is introduced to assist the phase transition. An unbroken $Z_2$ symmetry can ensure the inertness of $\chi$, keeping it disconnected from the rest of the phenomenology.

The tree level scalar potential of the SM-like Higgs doublet $H$ and neutrinophilic Higgs doublet $\eta$ is given by \cite{Davidson:2009ha}
\begin{align}
    V_0 & = -\mu_{1}^{2} |H|^{2}+\frac{\lambda_{1}}{2} |H|^{4} +m_{\eta}^2 |\eta|^2+\frac{\lambda_4}{2} |\eta|^4 \nonumber\\
&  + \lambda_5 (\eta^\dagger \eta) (H^\dagger H) + \lambda_6 (H^\dagger \eta)(\eta^\dagger H) \nonumber \\
   & + \frac{\lambda_7}{2} [ (H^\dagger \eta)^2 +{\rm h.c.} ] + (\mu^2_{\rm soft} \eta^\dagger H + {\rm h.c.})
\end{align}
where we consider the alignment limit for the two Higgs doublets. Due to the presence of soft $Z_2$-breaking terms in the scalar potential, the neutral component of $\eta$ can acquire a small VEV after electroweak symmetry breaking in order to generate a sub-eV Dirac neutrino mass while keeping the Dirac Yukawa coupling $y_{\alpha i}$ in the desired ballpark to realise leptogenesis at $T > T_{\rm sph}$.

The scalar potential of the scalar singlets is given by 
\begin{align}
    V(\sigma, \phi) & = - \frac{\mu^2_\sigma}{2} \sigma^2 -\frac{\mu^2_\phi}{2} \phi^2 + \frac{\lambda_\sigma}{4} \sigma^4 + \frac{\lambda_\phi}{4} \phi^4 \nonumber \\
    & + \frac{\lambda_{\sigma\phi}}{4} \sigma^2 \phi^2,
    \label{eq:scalpot}
\end{align}
While the temperature evolution of $\phi$ controls the LNV, the second singlet scalar $\sigma$ with a sizable coupling to $\phi$ assists in realizing the desired VEV profile of $\phi$ independent of electroweak symmetry breaking. To study the finite-temperature behavior of the potential, we include the Coleman-Weinberg correction $V_{\rm CW}$ \cite{Coleman:1973jx} together with the thermal correction $V_{\rm T}$ \cite{Dolan:1973qd,Quiros:1999jp}.


We consider a two-step phase transition such that the VEVs of $(\phi, \sigma)$ follow $(0,0) \rightarrow (0, v_\phi) \rightarrow (v_\sigma, 0)$. At a very high temperature $T>T_{c1}$, both the scalars lie in the symmetric phase $(0,0)$. The first transition occurs at $T=T_{c1}$, when the scalar $\phi$ attains a large VEV, and thereby generates the Majorana mass terms for $\nu_R$. The $\sigma$ field remains at zero during this time. As the temperature decreases, a FOPT occurs at $T=T_{c2}$. At this point, $\sigma$ gains a non-zero VEV while the $\phi$ VEV returns to zero which leads to purely Dirac neutrinos at low temperature. Subsequently, the VEV of $\sigma$ generates the mass for the dark fermion $\chi$ at present temperatures. The details of this two-step phase transition are given in Appendix \ref{appen1}.


\par

\begin{center}
    \begin{table}[ht]
\centering

\begin{tabular}{|c|c|c|c|}
\hline
 & BP 1 & BP 2 & BP 3 \\
\hline
$\lambda_\sigma$ 
& $0.6326$ 
& $0.7143$ 
& $0.6938$ \\
\hline
$\lambda_\phi$ 
& $0.1639$ 
& $0.2456$ 
& $0.2797$ \\
\hline
$\lambda_{\sigma\phi}$ 
& $0.6735$ 
& $0.8775$ 
& $0.9184$ \\
\hline
$v_\sigma$ [GeV]
& $1449.31$ 
& $1415.47$ 
& $1664.19$ \\
\hline
$v_\phi$ [GeV]
& $2000.0$ 
& $1817.67$ 
& $2056.22$ \\
\hline
$T_{c1}$ [GeV] 
& $2513.69$ 
& $2517.98$ 
& $2928.76$ \\
\hline
$T_{c2}$ [GeV] 
& $1340.03$ 
& $1355.38$ 
& $1498.01$ \\
\hline
$T_n$ [GeV] 
& $507.64$ 
& $635.75$ 
& $592.34$ \\
\hline
$M_{\nu_R}(T_n)$ [GeV] 
& $736.43$ 
& $667.62$ 
& $759.73$ \\
\hline
 $M_\phi (T_n)$ [GeV]& $1156.89$& $1277.98$&$1551.10$\\\hline
 $M_\sigma (T_n)$ [GeV]& $2290.40$& $2360.25$&$2752.90$\\\hline

$\beta/\mathcal{H}$ 
& $48.64$ 
& $856.78$ 
& $41.53$ \\
\hline
$\alpha$ 
& $0.0188$ 
& $0.00807$ 
& $0.0176$ \\
\hline
$y$ 
& $4.05\times10^{-4}$ 
& $7\times10^{-4}$ 
& $4.85\times10^{-4}$ \\
\hline
\end{tabular}
\caption{Benchmark points and corresponding phase transition parameters keeping RHN Yukawa ($Y_{\nu_R} = 0.37$) and dark fermion yukawa ($y_\chi = 0.87$) fixed. VEVs are at tree level without temperature correction while masses are quoted at the nucleation temperature $T_n$ using the full potential. Physical scalar masses are quoted at $T_n$ and field value of $(v_\sigma(T_n),0)$ at the true minimum of $(\sigma, \phi)$ potential.}
\label{tab:BM_points}
\end{table}
\end{center}

\begin{figure}[h]
    \centering
    \includegraphics[width=\linewidth]{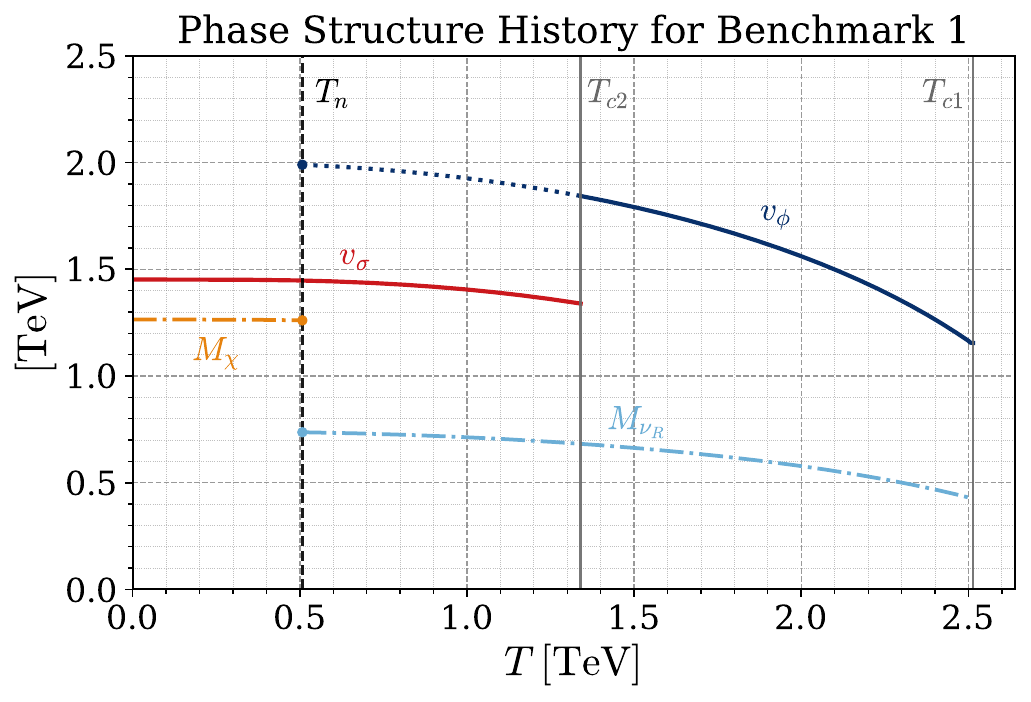}
    \caption{Thermal history of $(\sigma, \phi)$ phase for the benchmark point BP1 of Table \ref{tab:BM_points}. At $T< T_{c2}$, the $(0, v_\phi)$ VEV becomes metastable, as shown in the dotted line. At $T=T_n$, the Universe makes a transition into the true vacuum $(v_\sigma, 0)$.}
    \label{Fig:phaseBP1}
\end{figure}


\begin{figure*}
    \centering
    \includegraphics[scale=0.4]{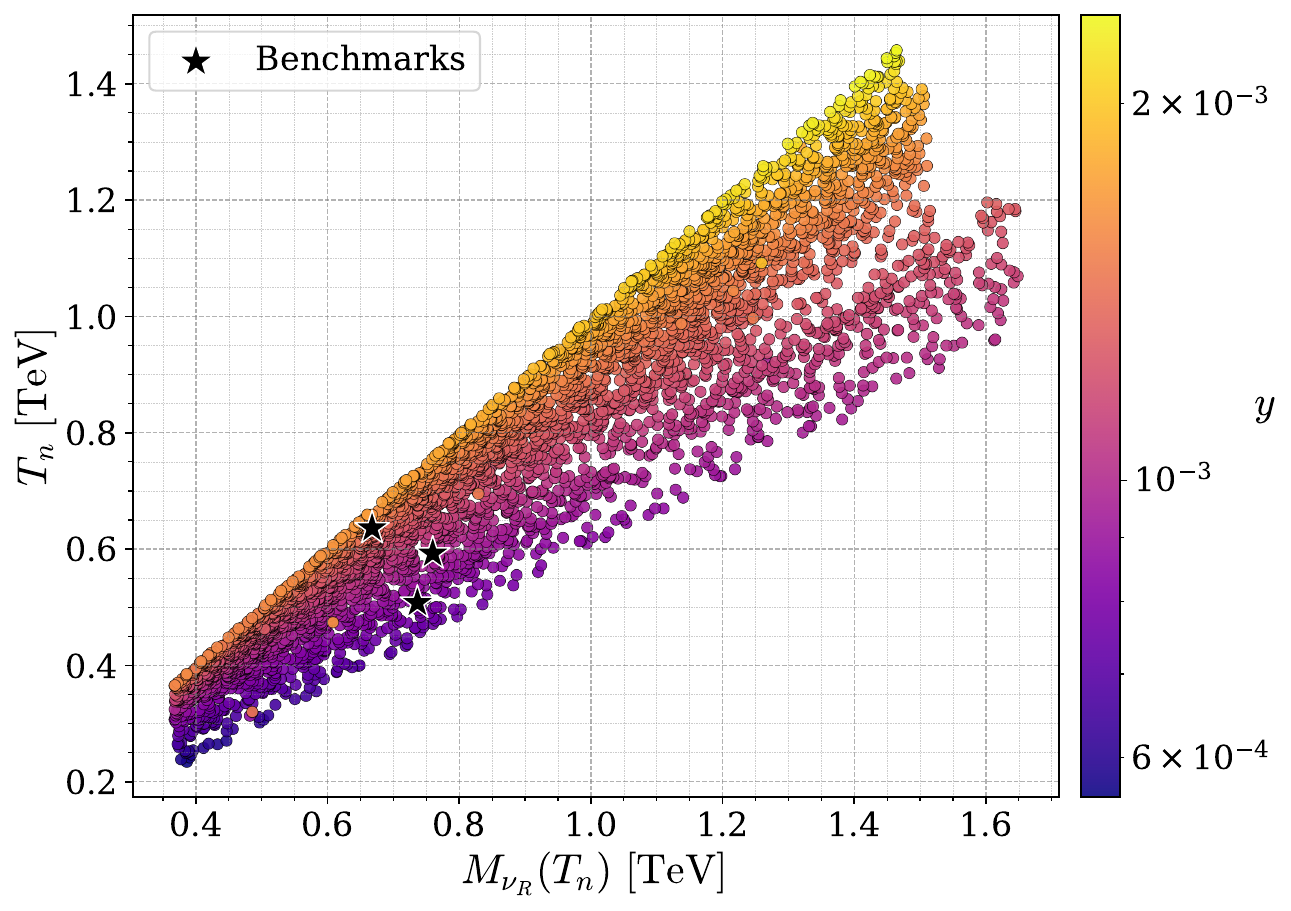}
     \includegraphics[scale=0.4]{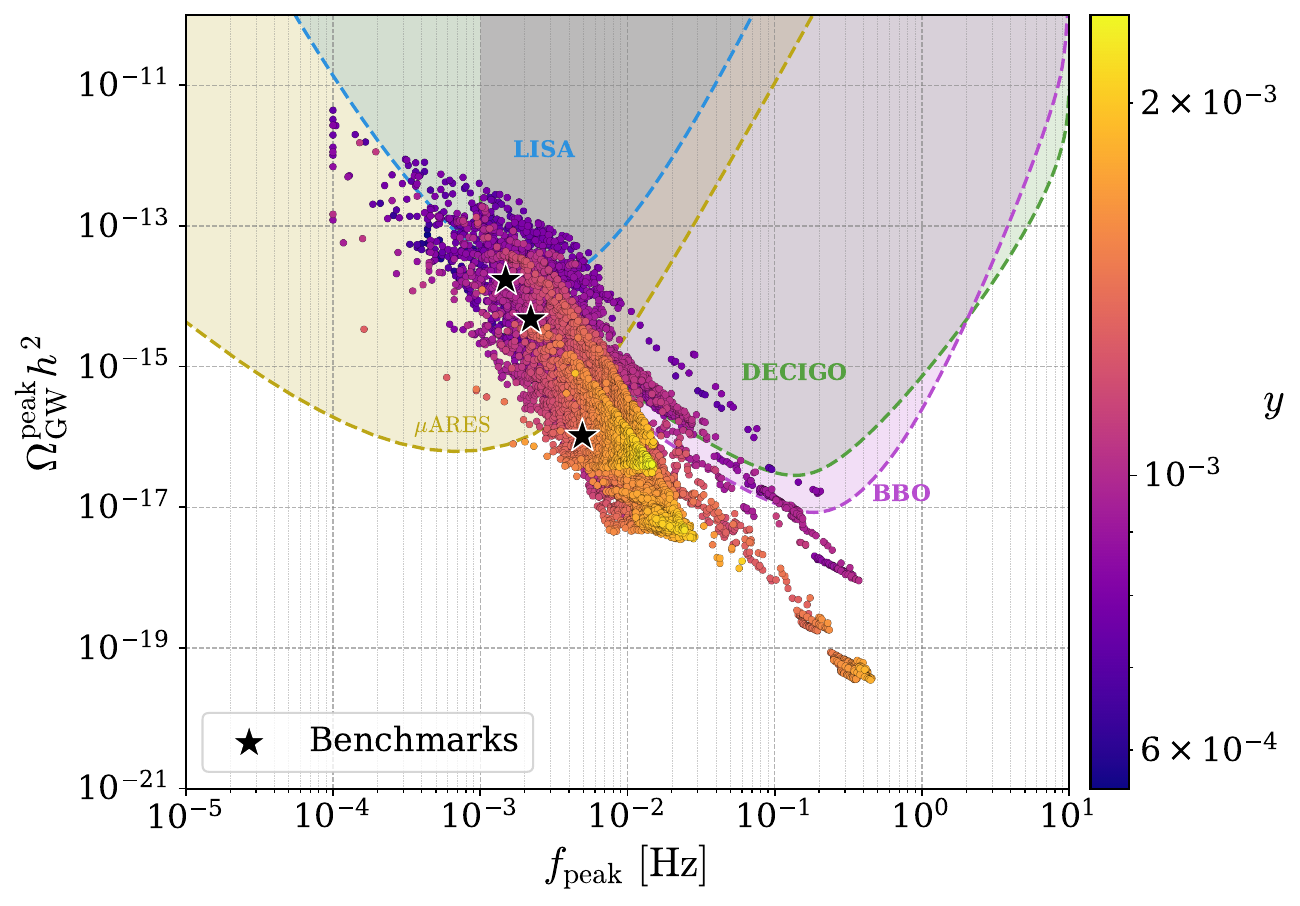}
    \caption{Left: Parameter space for successful leptogenesis and two-step FOPT in $T_n$-$M_{\nu_R}(T_n)=Y_{\nu_R} v_\phi(T_n)$ plane with Dirac Yukawa in color bar assuming CP asymmetry $\epsilon=0.1$ and strong washout regime. Right: Peak amplitudes of the gravitational wave spectra for the same parameter space on the left panel. The power-law sensitivity curves for the GW detectors are taken from \cite{Schmitz:2020syl}. On both the panels, the $\star$-shaped points correspond to the benchmark points in Table \ref{tab:BM_points}.}
    \label{Fig:GW1}
\end{figure*}

\begin{figure}
    \centering
    \includegraphics[scale=0.4]{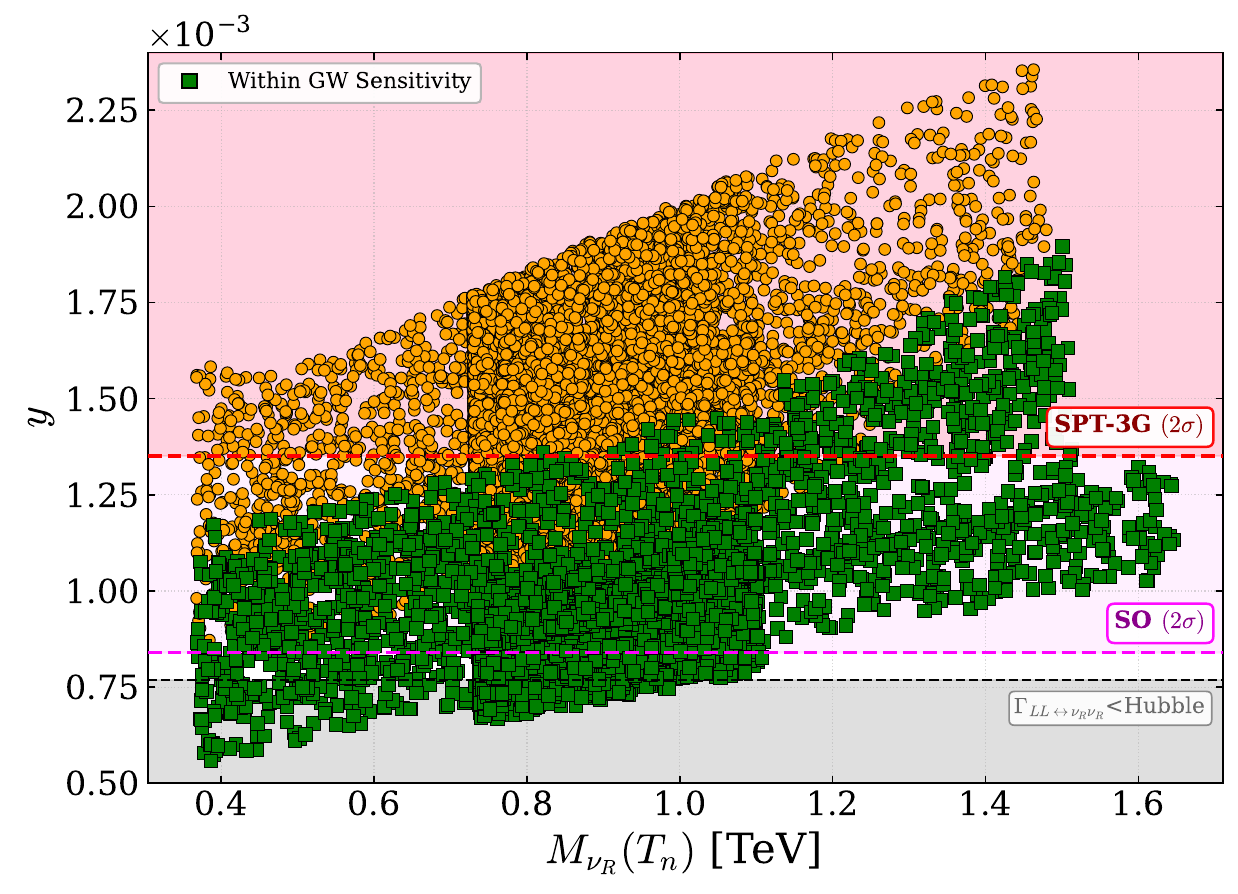}
    \caption{Same points as in Fig. \ref{Fig:GW1}, shown in $y-M_{\nu_R}(T_n)$ plane. The green points are within reach of future GW experiments. The shaded regions above the dashed contours are within reach of SPT-3G \cite{SPT-3G:2014dbx} and $\text{SO}$ \cite{SimonsObservatory:2018koc} respectively. The grey-shaded region below the black dashed contour keeps the process $L L \leftrightarrow \nu_R \nu_R$ out of equilibrium.}
    \label{Fig:summary_plot}
\end{figure}

\noindent 
\textbf{\textit{Results and Discussion:}} We first identify a few benchmark points given in Table \ref{tab:BM_points} which are consistent with a strong FOPT. While leptogenesis does not necessarily require a FOPT, this possibility naturally arises in our setup due to the existence of a strong potential barrier in the field space, particularly during the second step of the transition. The key parameters, namely, the latent heat released $(\alpha)$, inverse duration of the transition $(\beta/\mathcal{H})$, and the nucleation temperature $(T_n)$ for the chosen benchmark points are also given in Table \ref{tab:BM_points}. Fig. \ref{Fig:phaseBP1} shows the phases of this two-step FOPT for the benchmark point BP1 in Table \ref{tab:BM_points}. At the first critical temperature $T_{c1}$, the RHNs acquire masses due to the non-zero VEV of $\phi$. While the $\phi$ VEV drops to zero at the second critical temperature $T_{c2}$, the RHNs remain massive until the nucleation of bubbles at $T_n$. The lepton asymmetry can, therefore, be generated during $T_n < T < T_{c1}$. While thermal leptogenesis in a canonical seesaw framework with hierarchical RHNs is not possible below the Davidson--Ibarra bound $M_{\nu_R} \sim 10^9$ GeV~\cite{Davidson:2002qv}, the inclusion of a neutrinophilic Higgs doublet allows for such a possibility \cite{Haba:2011ra, Haba:2013pca, Clarke:2015hta}. For simplicity, we stick to the resonant leptogenesis regime \cite{Pilaftsis:2003gt} with quasi-degenerate RHNs. Assuming a maximal CP asymmetry $\epsilon \sim 0.1$, Table \ref{tab:BM_points} also lists the required Dirac Yukawa coupling $y$ needed to generate the required lepton asymmetry in the given temperature window $T_n < T < T_{c1}$. To estimate the lepton asymmetry, the relevant Boltzmann equations for comoving number densities $Y=n_X/s$ of $X \equiv \nu_R, B-L$ ($s$ being the entropy density) are solved numerically. In addition to considering the finite-temperature masses of $\nu_R, \eta, L$, we also consider the modified CP asymmetry parameter $\epsilon$ by appropriately considering such corrections. The washout terms for lepton asymmetry, like inverse decays and scatterings, are also considered in the Boltzmann equations. The details are given in Appendix \ref{appen2}. The lepton asymmetry at the sphaleron decoupling epoch $T_{\rm Sph} \sim 130$ GeV gets converted into baryon asymmetry as 
\begin{align}
& Y_B \simeq a_\text{sph}\,Y_{B-L} = \frac{1}{4} \,Y_{B-L}, 
\label{eq:sphaleron}
\end{align}
in agreement with earlier works \cite{Heeck:2013vha, Kartavtsev:2005rs}. The final baryon asymmetry $\eta_B$ can be analytically estimated to be \cite{Buchmuller:2004nz}
 \begin{align}
     \eta_B = \frac{a_{\rm sph}}{f} \epsilon \kappa\,, \label{eqn:etaana}
 \end{align}
where the factor $f$ accounts for the change in the relativistic degrees of freedom from the scale of leptogenesis until recombination and comes out to be $f=\frac{106.75}{3.91}\simeq27.3$. $\kappa$ is known as the efficiency factor, which incorporates the effects of washout processes. We numerically evaluate $Y_{B-L}$ at $T=T_{\rm sph}$ and use the above conversion factor to find the baryon asymmetry to fit the model predictions with the observed data. It is worth mentioning that for all the parameter space $T_n > T_{\rm sph}$, and it suffices to calculate $Y_{B-L}$ till $T=T_n$ below which lepton asymmetry saturates due to halt in production as well as lack of washout due to the absence of LNV.

After identifying a few benchmark points consistent with the requirements of successful leptogenesis and two-step FOPT, we perform a numerical scan of the parameter space satisfying these criteria. For the scan, we keep these Yukawa couplings of $\nu_R, \chi$ to singlet scalars fixed as $Y=0.37$ and $Y_\chi = 0.87$. The left panel of Fig. \ref{Fig:GW1} shows the parameter space in the plane of nucleation temperature $T_n$ and the scale of leptogenesis $M_{\nu_R}(T_n)$ assuming $\epsilon=0.1$. The color code indicates the required Dirac Yukawa coupling for leptogenesis. The size of these Dirac Yukawa couplings keep the leptogenesis in the strong washout regime. For a fixed scale of leptogenesis, we need a larger Yukawa coupling for a larger $T_n$ to generate the same lepton asymmetry. This is due to requirement of producing the same asymmetry in a smaller temperature window $T_n < T < T_{c1}$. The benchmark points of Table \ref{tab:BM_points} are also highlighted as $\star$-shaped points in the figure. The right panel of Fig. \ref{Fig:GW1} shows the peak amplitudes of the GW spectrum generated during the FOPT from from bubble collisions \cite{Turner:1990rc,Kosowsky:1991ua,Kosowsky:1992rz,Kosowsky:1992vn,Turner:1992tz}, the sound wave of the plasma~\cite{Hindmarsh:2013xza,Giblin:2014qia,Hindmarsh:2015qta,Hindmarsh:2017gnf} and the turbulence of the plasma~\cite{Kamionkowski:1993fg,Kosowsky:2001xp,Caprini:2006jb,Gogoberidze:2007an,Caprini:2009yp,Niksa:2018ofa}. The shaded regions show the sensitivity of LISA~\cite{2017arXiv170200786A}, BBO~\cite{Yagi:2011wg}, DECIGO~\cite{Seto:2001qf,Kawamura:2006up,Yagi:2011wg} and $\mu$ARES \cite{Sesana:2019vho}.  \\

Fig. \ref{Fig:summary_plot} shows a summary of the parameter space in the plane of Dirac neutrino Yukawa $y$ and the scale of leptogenesis $M_{\nu_R}(T_n)$. The green colored points remain within reach of one of the above-mentioned GW detectors. The shaded regions remain within reach of South Pole Telescope (SPT-3G) \cite{SPT-3G:2014dbx} and Simons Observatory (SO) \cite{SimonsObservatory:2018koc} capable of doing high-resolution CMB surveys. As thermalized Dirac neutrinos lead to enhanced dark radiation, precision CMB measurements can probe such a scenario \cite{Abazajian:2019oqj, Luo:2020sho}. The grey-shaded region keeps the interaction rate $\Gamma (L L \leftrightarrow \nu_R \nu_R)$ below the Hubble expansion rate\footnote{Such small Yukawas can still lead to non-thermal production of dark radiation \cite{Luo:2020fdt, Biswas:2022vkq} which we do not pursue here.}. Clearly, the parameter space consistent with successful leptogenesis and a two-step FOPT has complementary detection prospects in terms of CMB surveys and GW detectors. \\

\noindent 
\textbf{\textit{Conclusion:}} We have proposed a novel leptogenesis scenario by considering the possibility of neutrinos doing a flip-flop in their nature in the early Universe due to finite-temperature effects. While light neutrinos are of Dirac type, consistent with the absence of LNV signatures like neutrinoless double beta decay, their right chiral parts acquire a large Majorana mass in the early Universe due to a singlet scalar field acquiring a non-zero VEV. This allows the generation of lepton asymmetry in the conventional way, which later gets converted into baryon asymmetry via electroweak sphalerons. The VEV of the scalar field gets restored to zero at a lower temperature to retain the pure Dirac nature of light neutrinos after electroweak symmetry breaking. The second step of the transition can naturally be of first-order resulting in additional detection probe via future observations of stochastic gravitational waves at experiments like LISA, BBO and DECIGO. Presence of a neutrinophilic Higgs doublet allow sizeable Yukawa coupling of Dirac neutrinos, assisting in leptogenesis as well as thermalizing their right chiral parts. This also leads to enhance dark radiation, keeping the model within reach of CMB experiments like SPT-3G and Simons Observatory. The neutrinophilic Higgs doublet and the two singlet scalars can have complementary detection prospects at colliders. While we have implemented our idea in a toy model, it can open up new directions in UV complete model building with additional symmetries, fields as well as related phenomenology, which are left for future works.

\acknowledgements
The work of D.B. and I.S. are supported by the Science and Engineering Research Board (SERB), Government of India grant CRG/2022/000603.

\appendix

\section{Two-step Phase Transition}
\label{appen1}

To study the dynamics of the phase transition, we calculate the effective potential at finite temperature as
\begin{align*}
    V_{\text{eff}}(\sigma,\phi,T) = &V(\sigma, \phi) + V_{CW}(\sigma, \phi ) + V_{\rm T}(\sigma, \phi, T) \\
    & + V_{\text{Daisy}}(\sigma, \phi, T)
\end{align*}
where, $V_{CW}(\sigma, \phi )$ is the zero temperature Coleman-Weinberg potential~\cite{Coleman:1973jx}. It is given by,
\begin{equation}
\label{one-loop}
    V_{\text{CW}}= \sum_{i=\sigma,\phi,\nu_{R_i}, \chi}  (-1)^f\frac{n_i m_i^4(\sigma,\phi)}{64\pi^2}\Biggl[\ln\Bigl(\frac{m_i^2(\sigma,\phi)}{\mu^2}\Bigr)-C_i\Biggr]
\end{equation}
where $C_i=3/2$, $n_i=\{1,1,2,2\}$ and $f=0(1)$ for scalars (fermions). The zero-temperature field-dependent masses are given by,
\begin{align*}
    m_\sigma^2(\sigma,\phi) &= -\mu_\sigma^2 + 3\lambda_\sigma\sigma^2 + \frac{\lambda_{\sigma\phi}}{2}\phi^2 \\
    m_\phi^2(\sigma,\phi) &= -\mu_\phi^2 + 3\lambda_\phi\phi^2 + \frac{\lambda_{\sigma\phi}}{2}\sigma^2 \\
    m_{\nu_{R_i}}^2 (\phi) &= \sum_{i,j}|Y_{ij}|^2\phi^2 \\
    m_{\chi}^2(\sigma) &= Y_{\chi}^2\sigma^2
\end{align*}
The thermal correction to the tree-level potential can be written as~\cite{Dolan:1973qd,Quiros:1999jp} ,
\begin{equation}
    V_T = \frac{T^4}{2\pi^2}\Biggl[ \sum_{b=\sigma,\phi}n_bJ_B\Bigl(\frac{m_b^2}{T^2}\Bigr)-\sum_{f=\nu_{R_i},\chi}n_fJ_F\Bigl(\frac{m_f^2}{T^2}\Bigr)\Biggr]
\end{equation}
where, $n_\sigma, n_\phi = 1$ and $n_{\nu_{R_i}} = n_{\chi}=2$, $(i = 1,2)$. The bosonic and fermionic thermal functions are
\begin{equation}
    J_{B/F}(x^2)
\equiv
\int_{0}^{\infty} dy\, y^2
\ln\left[1 \mp e^{-\sqrt{y^2+x^2}}\right]
\end{equation}


 To maintain a consistent perturbative expansion in the high temperature limit, we add the Daisy corrections. In the Arnold-Espinosa prescription \cite{Arnold:1992rz}, the resummation of Daisy diagrams gives the Daisy contribution $V_{\rm Daisy}$ to $V_{\rm eff}$, given by,
 \begin{equation*}
\label{daisy}
    V_{\rm Daisy}=-\frac{T}{12\pi}\sum_{b=\sigma,\phi}g_b\Bigl[\bigl(m_b^2(x_a)+\Pi_b(T)\bigr)^{3/2}- m_b^3(x_a)\Bigr]
\end{equation*}
To leading order, the bosonic thermal masses $\Pi_b$ are field independent, and are given as,
\begin{align}
    \Pi_\sigma^2(T) &= \frac{1}{24}\bigl(\lambda_{\sigma\phi} + 6\lambda_\phi + 2\sum_{i,j}|Y_{i,j}|^2\bigr)T^2, \\
    \Pi_\phi^2(T) &= \frac{1}{24}\bigl(\lambda_{\sigma\phi} + 6\lambda_\sigma + 2|Y_{\chi}|^2\bigr)T^2
\end{align}

The tunneling rate per unit volume is determined by the O(3)-symmetric bounce action, $S_3(T)$ and can be expressed as $\Gamma(T)\simeq T^4\exp\left[-\frac{S_3(T)}{T}\right]$. The nucleation temperature $T_n$ is then obtained by requiring the tunneling rate to become comparable to the Hubble expansion rate, namely, $\Gamma(T_n)=\mathcal{H}^4(T_n)$. The strength of the phase transition is characterized by the dimensionless parameter $\alpha \equiv \frac{\epsilon}{\rho_{\rm rad}(T_n)}$, where $\epsilon = \left(\Delta V_{\rm eff} - \frac{T}{4}\frac{\partial \Delta V_{\rm eff}}{\partial T}\right)|_{T=T_n}$ denotes the released vacuum energy density and $\rho_{\rm rad}(T) = \frac{\pi^2}{30}g_*T^4$ is the radiation energy density. The inverse duration of the phase transition is quantified by the dimensionless parameter $\beta/\mathcal{H}$, given by
$\beta/\mathcal{H} \equiv T\left.\frac{d}{dT}\left(\frac{S_3(T)}{T}\right)\right|_{T=T_n}$. We implement this full effective potential in the \texttt{CosmoTransitions} \cite{Wainwright:2011kj} package to estimate bounce action along with parameters associated with gravitational waves. To calculate the velocity of bubble wall, we follow the prescription~\cite{Lewicki:2021pgr},
\begin{equation}
    v_w =
\begin{cases}
\displaystyle
\sqrt{\frac{\Delta V_{\rm eff}}
{\alpha \rho_{\rm rad}}},
& \displaystyle
\sqrt{\frac{\Delta V_{\rm eff}}
{\alpha \rho_{\rm rad}}}<v_J,
\\[10pt]
1,
& \displaystyle
\sqrt{\frac{\Delta V_{\rm eff}}
{\alpha \rho_{\rm rad}}}\geq v_J,
\end{cases}
\end{equation}
where $v_J$ is Jouguet velocity~\cite{Espinosa:2010hh}. For the region of our interest, we have $v_w=1$.

\section{Details of Leptogenesis}
\label{appen2}
The CP asymmetry parameter corresponding to the CP-violating decay of massive RHN $\nu_R$ (summing over all lepton flavours) is given by \cite{Pilaftsis:2003gt}
\begin{equation}\begin{aligned}
\epsilon_{i} & = \dfrac{\Gamma(\nu_{R_i}\to\sum_{\alpha}L_{\alpha} \eta )-\Gamma(\nu_{R_i}\to \sum_{\alpha}\bar L_{\alpha}\bar{\eta })}{\Gamma(\nu_{R_i}\to \sum_{\alpha}L_{\alpha} \eta)+\Gamma(\nu_{R_i}\to\sum_{i}\bar L_{\alpha}\bar{\eta})}, \label{eq:asymmparameter}
 \end{aligned}\end{equation}
which, after including finite-temperature effects, leads to \cite{Borah:2023god}
\begin{widetext}
\begin{equation}
    \epsilon_i =\left[\left(M_{\nu_{R_i}}^2 + m_{L}^2 - m_\eta^2\right)\lambda^{1/2}\left(M_{\nu_{R_i}}^2,m_{L}^2,m_\eta^2\right)\Theta\left(M_{\nu_{R_i}}^2 - \left(m_\eta + m_{L}\right)^2\right)\right]\frac{1}{(y^{\dagger}y)_{ii}}\sum_{j\neq i}\frac{{\rm Im}[((y^{\dagger}y)_{ij})^{2}]}{16\pi M_{\nu_{R_i}}^3} \frac{M_{\nu_{R_j}}\Delta_{ij}}{\Delta_{ij}^2 + \left(M_{\nu_{R_j}}\Gamma_{N_j}\right)^2} 
\end{equation}
\end{widetext}
where $\Delta_{ij} \equiv M_{\nu_{R_i}}^2 - M_{\nu_{R_j}}^2, \lambda(x,y,z) \equiv x^2 + y^2 + z^2 - 2xy - 2xz - 2yz $ and $\Theta(x)$ is the Heaviside step function. It should be noted that in the above derivation of temperature corrected CP asymmetry, we consider the interference of tree level and self-energy diagrams only, assuming a quasi-degenerate RHN spectrum for which the vertex diagram contribution remains sub-dominant.

In the following, we furnish the details of the thermal masses of the neutrinophilic Higgs doublet, used in the leptogenesis calculation.

\begin{equation}
     \eta=\begin{pmatrix}
        \eta^{\pm}\\
        \frac{\eta_0+iA}{\sqrt{2}}
    \end{pmatrix}
\end{equation}
The thermal masses of it's components are given by
\begin{align}
    m_{\eta^{\pm}}^2(h,T) &= (\lambda_5/2)h^2 +m_{\eta}^2 + \Pi_{\eta}(T),\\
    m_{\eta_0}^2(h,T) &= \frac{\lambda_{\eta_{0}}}{2}h^2+m_{\eta}^2 + \Pi_{\eta}(T),\\
    m_{A}^2(h,T) &= \frac{\lambda_A}{2}h^2+m_{\eta}^2 + \Pi_{\eta}(T),\\
\end{align}
where, $\lambda_{\eta_0,A}=\lambda_5 + \lambda_6 \pm \lambda_7$ and
\begin{align}
    \Pi_{\eta}(T)= \frac{1}{8}g^2T^2+\frac{1}{16}(g^2+g^{'2})T^2+\frac{1}{4}\lambda_4 T^2+\frac{1}{12}\lambda_5 T^2 \nonumber\\
    +\frac{1}{24}(\lambda_{\eta_0}+\lambda_A) T^2.
\end{align}
Thermal mass for the lepton is given by, 
\begin{equation}
    m_L^2(T)= \Biggl(\frac{3}{32}g^2 + \frac{1}{32}g'^2\Biggr)T^2.
\end{equation}
The relevant Boltzmann equations are as follows,

\begin{align}
\frac{dY_{\nu_{R_i}}}{dz} & = -D_{i}\left(Y_{\nu_{R_i}}-Y_{\nu_{R_i}}^{\rm eq}\right),  \label{eq:nu}\\
\frac{dY_{B-L}}{dz} &= -\sum_{i}\epsilon_{i}D_{i}\left(Y_{\nu_{R_i}}-Y_{\nu_{R_i}}^{\rm eq}\right) \nonumber \\
&-\left(\Delta W+\sum_{i}W_{i}\right)Y_{B-L}, \label{eq:B-L}
\end{align}
where, $Y_{\nu_{R_i}}(Y_{B-L})$ denote the comoving number densities of $\nu_{R_i}(B-L)$, and the index $i$ runs only over 1 and 2. While there exist annihilation channels of $\nu_R$, they remain kinematically forbidden for the range of our interest. Therefore, we only retain the decay and inverse decay contributions on the right hand side of Eq. \eqref{eq:nu}.
\begin{figure}
    \centering
    \includegraphics[scale=0.45]{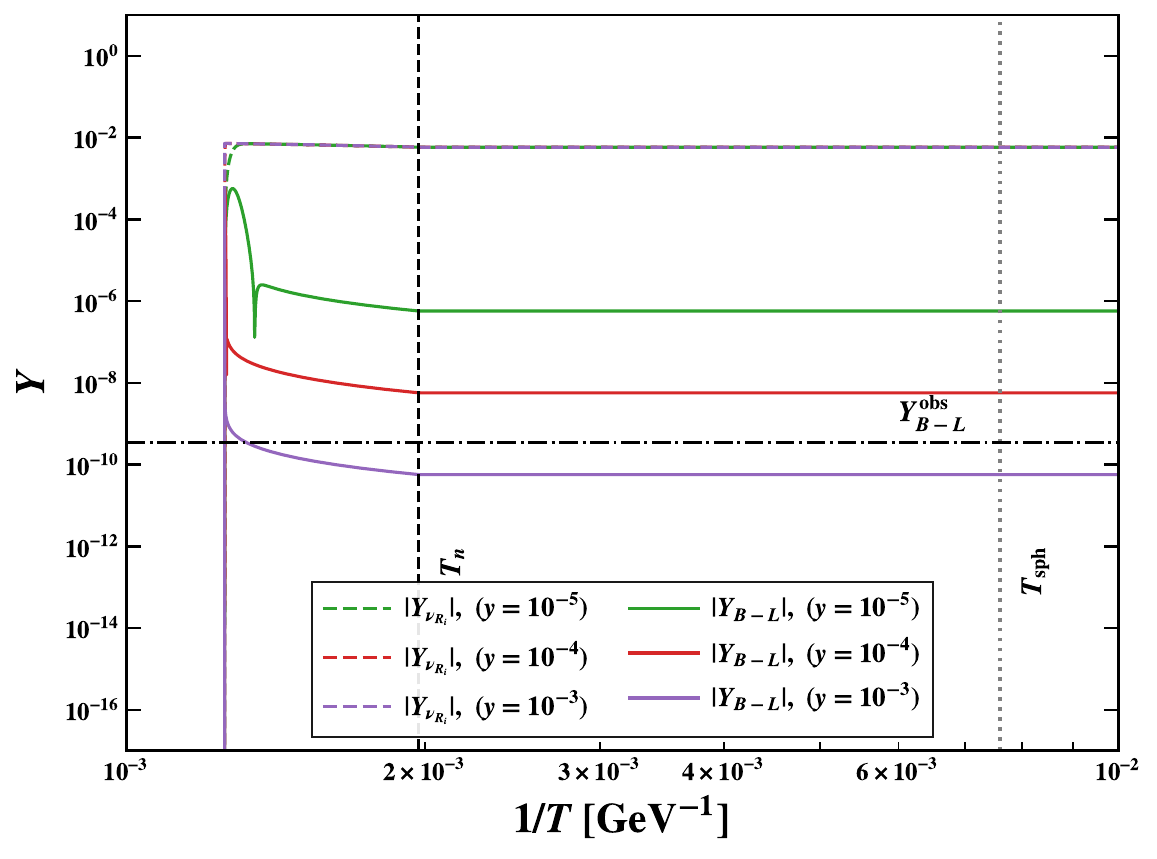}
    \caption{Evolution of comoving abundances for BP1 for fixed CP asymmetry $\epsilon=0.1$ and varying Dirac Yukawa coupling $y$.}
    \label{Fig:Lepto}
\end{figure}
The equilibrium no densities of $\nu_{R_i}$'s are defined by $Y_{\nu_{R_i}}^{\rm eq}=\frac{z^2}{2}\kappa_2(z)$, (with $\kappa_i(z)$ being the modified Bessel function of $i$-th kind) and $z=M_{\nu_{R_1}}/T=M_{\nu_{R_2}}/T$ (for resonant leptogenesis). $D_{1,2}$ are the decay terms and $W_{1,2}$ are the inverse decay terms for $\nu_{R_1}$ and $\nu_{R_2}$ decays, respectively, which are measures of the rate of decay and inverse decay with respect to the background expansion of the universe parametrized by Hubble expansion rate $\mathcal{H}$. They are defined as

\begin{eqnarray}
D_{i} & = & \dfrac{ \langle \Gamma_{i} \rangle}{{\mathcal{H}} z} = K_{i}z\dfrac{\kappa_{1}(z)}{\kappa_{2}(z)},  \\
W_{i} & = & \dfrac{1}{4}K_{i}z^{3}\kappa_{1}(z),
\end{eqnarray}
with $K_i=\Gamma_i/{\mathcal{H}}(z=1)$ being the decay parameter. The term $\Delta W$ on the right-hand side of the equation \eqref{eq:B-L} takes account of all the scattering processes that can act as possible washouts for the generated $B-L$ asymmetry. The important washout processes are
\begin{align*}
    \Delta L&=1: \quad L W^{\pm} (Z)\longrightarrow \nu_R \eta,\, L \eta \longrightarrow \nu_R W^{\pm} (Z)\\
    \Delta L&=2: \quad L \eta \longrightarrow L^{c}\eta^*.
\end{align*}
The resulting washout term is defined as,
\begin{equation}
    \Delta W=\dfrac{\langle \Gamma_{\rm scatterings} \rangle}{{\mathcal{H}} z}=\dfrac{1}{ z{\mathcal{H}}\, n_L^{\rm eq}}\left(\gamma_{\Delta L=1} +2\gamma_{\Delta L=2}\right)
\end{equation}
where
$\gamma$ is the thermally-averaged reaction density of a two-body scattering and is defined as ,
\begin{equation}
    \gamma_{ab\longleftrightarrow ij}=\frac{T}
    {64\pi^4}\int_{\rm s_{min}}^{\infty} ds\, s^{1/2}\,\hat{\sigma}(s)\,K_1\left(\frac{\sqrt{s}}{T}\right),
\end{equation}
where $s$ is the squared center of mass energy, ${\rm s_{min}= max}[(m_a+m_b)^2,(m_i+m_j)^2]$. Here $\hat{\sigma}\equiv 2s\, \lambda[1,m_a^2/s,m_b^2/s]\,\sigma$ is the reduced cross section, and $\sigma$ is the total cross section summed over final and initial quantum numbers. We numerically solve Eqs.~\eqref{eq:nu} and \eqref{eq:B-L} to study the evolution of the lepton asymmetry for different values of the Yukawa coupling $y$ associated with the decay of $\nu_R$. The corresponding results are shown in Fig.~\ref{Fig:Lepto} for BP1 in Table \ref{tab:BM_points} with fixed CP asymmetry $\epsilon=0.1$ but varying Dirac Yukawa coupling $y$. As the Yukawa coupling increases, washout due to inverse decay increases leading to smaller lepton asymmetry.


\providecommand{\href}[2]{#2}\begingroup\raggedright\endgroup

\end{document}